\pdfoutput=1

\documentclass[]{spie}  
\usepackage[]{graphicx}

\newcommand{\A}{{A}}
\newcommand{\Ad}{{A^\dag}}
\newcommand{\Sa}{{S}}
\newcommand{\Sd}{{S^\dag}}

\newcommand{\Fd}{{F^\dag}}
\newcommand{\W}{{W}}
\newcommand{\Wd}{{W^\dag}}

\newcommand{\dnuno}{{\left(\frac{\nu-\nu_0}{\nu_0}\right)}}

\newcommand{\dtmto}{{\left(\frac{2(t-t_{mid})}{t_{max}-t_{min}}\right)}}

\newcommand{\V}{\vec{V}}
\newcommand{\I}{\vec{I}}

\title{Radio interferometric imaging of spatial structure that varies with time and frequency} 

\author{Urvashi Rau \skiplinehalf
National Radio Astronomy Observatory, 1003 Lopezville Road, Socorro, NM-87801, USA \\
}

\authorinfo{E-mail: rurvashi@aoc.nrao.edu, Telephone: 1 575 835 7372}

\begin{document} 
  \maketitle 

\begin{abstract}
The spatial-frequency coverage of a radio interferometer is increased by combining samples acquired at different times and observing frequencies.  However, astrophysical sources often contain complicated spatial structure that varies within the time-range of an observation, or the bandwidth of the receiver being used, or both.  Image reconstruction algorithms can been designed to model time and frequency variability in addition to the average intensity distribution, and provide an improvement over traditional methods that ignore all variability. This paper describes an algorithm designed for such structures, and evaluates it in the context of reconstructing three-dimensional time-varying structures in the solar corona from radio interferometric measurements between 5 GHz and 15 GHz using existing telescopes such as the EVLA and at angular resolutions better than that allowed by traditional multi-frequency analysis algorithms.
\end{abstract}



\section{INTRODUCTION}
\label{sec:intro}  

Image reconstruction in radio interferometry\cite{NRAO_LECTURES} benefits from maximizing the number of distinct spatial frequencies at which the visibility function of the target is sampled.  For a given array, the spatial-frequency coverage can be increased by taking measurements across several hours (Earth-rotation-synthesis) and at multiple observing frequencies (multi-frequency-synthesis).  However, astrophysical sources sometimes show time and frequency variability within the parameters of such an observation.  Reconstruction algorithms have traditionally ignored this variability, leading either to imaging-artifacts or to data-analysis strategies that use subsets of the data independently, both which limit imaging sensitivity and accuracy.

In recent years, broad-band receivers have been installed on radio interferometers specifically to increase imaging sensitivity, and this has necessitated the development of algorithms\cite{MFCLEAN_CCW, MFCLEAN, MSMFS2011} that account for sky spectra in addition to average intensity.
A similar idea applies to time-variability as well, and recent work\cite{MBCLEAN2011} has demonstrated combined time and frequency modeling for point-sources (without multi-scale support), mainly for the purpose of eliminating artifacts from images of the average intensity distribution.  

This paper presents a reworking of the system of equations being solved when modeling time and frequency varying structure, and includes multi-scale support. Using a software implementation of a multi-scale, multi-frequency, time-variable image-reconstruction algorithm that can be operated in modes that emulate various intermediate algorithms, a series of imaging tests were carried out on data simulated for three-dimensional time-varying structure in the solar corona. The results are discussed in terms of the accuracy of reconstructing the time and frequency-dependent structure at an angular resolution better than what traditional methods allow. 

\section{Model-fitting to reconstruct an image} 
\label{sec:algorithms}


According to the van-Cittert-Zernicke (VCZ) theorem\cite{NRAO_LECTURES} , an ideal radio interferometer samples the visibility function of the sky brightness distribution.
For a single frequency channel $\nu$ and integration timestep $t$, each pair of detectors corresponds to one visibility sample, located at the spatial-frequency defined by the distance between the pair, the phase-tracking center, and the observing frequency. With $n_{ant}$ antennas, $n_{ant}(n_{ant}1)/2$ such measurements of the visibility function are made at spatial frequencies that define the $uv$-coverage of the instrument. Image-reconstruction is then a non-linear process that solves for the parameters of a sky model, subject to constraints from the data measured at this set of spatial frequencies.
The instantaneous single-channel $uv$-coverage is usually augmented by Earth-rotation synthesis and multi-frequency synthesis.  Under the assumption that the sky structure is invariant with time and frequency, these samples can be considered additional independent measurements of the same visibility function. The VCZ theorem holds, and standard image-reconstruction algorithms such as CLEAN\cite{CLARK_CLEAN, CLEAN} and MEM\cite{MEM, MEM_RN}
apply, and benefit from the extra data-constraints. 

When the sky structure varies with frequency or time, within the parameters of a single observation, the traditional solution has been to partition the data in time and frequency, such that the VCZ theorem holds on each piece. This approach sacrifices sensitivity and $uv$-coverage during the non-linear reconstruction process. While this suffices for simple point-like structures and emission with high signal-to-noise ratios, non-uniqueness in the reconstruction of complicated emission can introduce errors that are inconsistent across time and/or frequency.   In such cases, it is better to model and fit for the frequency and time-dependence at the same time as the average intensity\cite{MBCLEAN2011}. Although this approach has primarily been applied to separate the varying and non-varying components of the dataset with the goal of producing a single image representing non-variant structure, it is sometimes possible to reconstruct the time and/or frequency variation with enough accuracy to be used for astrophysical interpretation. Such an analysis is strongly dependent on the models used during reconstruction, how well the measurements constrain their parameters, and how well the reconstruction algorithm is able to converge to the best-fit solution. 


For several decades, the general approach to image reconstruction in radio interferometry has been to describe the sky structure as a collection of delta-functions, parameterized by their location and amplitude\cite{CLEAN}.  For sky emission comprised of several discrete sources, delta-functions are a compact representation of the signal being modeled.  
More recently, multi-scale algorithms\cite{MSCLEAN, Asp_Clean, SPARSERI2010, LICS2011, SARA2012} have been used to model extended emission, by decomposing the sky brightness into a linear combination of basis functions representing flux-components of different shapes and sizes. This is considered a sparse representation of extended spatial structure, and the parameters to be solved-for are the location, amplitude, and size of the components.
Multi-frequency algorithms\cite{MFCLEAN_CCW, MFCLEAN, MSMFS2011} model and reconstruct the sky spectrum by solving for the coefficients of a Taylor-polynomial representation of the spectrum, again, a sparse representation of smooth spectral structure. This idea can be extended to time-variability\cite{MBCLEAN2011} as well, in which a set of basis functions is chosen so as to compactly represent the time-variation. 

In most of these cases, the image-reconstruction algorithm first transforms the data into a basis in which the model has a sparse representation, and then applies a greedy algorithm to search for flux components and their best-fit parameters.  

MS-MF-TV-CLEAN, an  algorithm that uses a sparse representation of multi-scale, frequency-dependent, time-variable structure and solves for its parameters via $\chi^2$ minimization is described below.  Differences with the multi-frequency and time-variable algorithms in Refs.~\citenum{MSCLEAN, MFCLEAN,MBCLEAN2011} and the consequences of these differences will be pointed out when relevant.  Sec.~\ref{sec:examples} uses some imaging results to evaluate the accuracy at which such methods are able to reconstruct time and frequency-variable structure, using $uv$-coverages offered by moderately sized interferometers such as the EVLA with 27 antennas and a scaled-down FASR with 15 antennas.

\subsection{Multi-term image models}
Parameterized models for multi-scale, frequency-dependent and time-varying spatial structure, and their combination as used in the MS-MF-TV-CLEAN algorithm are described below.

\paragraph{Multi-Scale structure :} 
For a finite set of $N_s$ spatial scales, a multi-scale image model is written as
\begin{equation}
\vec{I}^{m} = \sum_{s=0}^{N_s-1}  \vec{I}^{shp}_{s} \star \vec{I}^{sky}_s
\label{Eq:ms_model}
\end{equation}
where $\vec{I}^{shp}_s$ are multi-scale basis functions, and 
$\vec{I}^{sky}_{s}$ is an image of delta-functions that mark the locations and amplitudes
of the basis function of type $s$. 
As implemented in MS-CLEAN\cite{MSCLEAN} , the first scale 
function $\vec{I}^{shp}_{s=0}$ is chosen to be a $\delta$-function in order to always allow 
for the modeling of unresolved sources, and successive basis functions are
inverted and truncated parabolas of increasing width (as $s$ increases).

\paragraph{Multi-Frequency structure} 
A sky brightness distribution that varies smoothly with observing frequency can be modeled
with Taylor-polynomials per pixel, and represented as a linear combination of 
coefficient images and basis functions\cite{MFCLEAN, MSMFS2011, MBCLEAN2011}. 
The intensity at each frequency channel can be written as
\begin{eqnarray}
\label{Eq:freqfunc}
\vec{I}^{m}_{\nu} = \sum_{p=0}^{N_p-1} \vec{I}^{sky}_{p} \dnuno^p
\end{eqnarray}
where $N_p$ is the total number of terms in the series 
and $\vec{I}^{sky}_{p} ~ ; 0\le p<N_p$ are the coefficient images (one set of coefficients per pixel).
$\nu$ is the observing frequency, $\nu_0$ is a chosen reference frequency.
%

\paragraph{Time-Variable structure} 
A smooth variation of the source intensity as a function of time can be described as another Taylor-polynomial.
\begin{eqnarray}
\label{Eq:timefunc}
\vec{I}^{m}_{t} = \sum_{q=0}^{N_q-1} \vec{I}^{sky}_{q} \dtmto^q
\end{eqnarray}
where $N_q$ is the total number of terms in the series 
and $\vec{I}^{sky}_{q}~ ; 0\le q<N_q$ are the coefficient images (one set of coefficients per pixel).
$t$ is the time, and $t_{mid}, t_{max}, t_{min}$ are calculated from the time-range
of the observation. These basis functions correspond to a Taylor polynomial evaluated
between -1 and +1.  Ref.~\citenum{MBCLEAN2011} discusses and demonstrates that if the time variation is not smooth, a Fourier-basis is much more appropriate, but requires many more terms in the series.

\paragraph{Multi-scale multi-frequency time-variable structure}
The basis functions described in Eqns.~(\ref{Eq:ms_model, Eq:freqfunc}) and (\ref{Eq:timefunc}) 
can be combined as follows.
\begin{eqnarray}
\label{Eq:combfunc}
\vec{I}^{m}_{\nu,t} = \sum_{s=0}^{N_s-1}  \vec{I}^{shp}_{s} \star \left[
 \vec{I}^{sky}_{0} + \sum_{p=1}^{N_p-1} \vec{I}^{sky}_{{p}} \dnuno^{p} +  \sum_{q=1}^{N_q-1} \vec{I}^{sky}_{q+N_p-1} \dtmto^q  \right ]_s
\end{eqnarray}
where the expression inside the square brackets represents images of delta functions that mark the
locations of flux components of shape $\vec{I}^{shp}_s$ and 
whose amplitudes follow dynamic-spectra defined by 
linear combinations of polynomials along the time and frequency axes.

With $N_s$ multi-scale basis functions, $N_p$ Taylor-terms along frequency, and $N_q$ Taylor-terms along time, the total number of basis functions becomes $N_s \times (N_p + N_q - 1)$.  
Alternately, one could use Zernicke polynomials to describe patterns on the 2D
time-frequency plane. 

Note that for each scale-size, this choice of basis-set requires fewer terms than the $N_p \times N_q$ basis functions used in the multi-beam clean algorithm described in Ref.~\citenum{MBCLEAN2011} which also uses a Gram-Schmidt orthogonalization to regularize the problem.  Also, the algorithm described here includes a multi-scale model, which from the analysis done to develop the MS-MF-CLEAN\cite{MSMFS2011} becomes very relevant when attempting to use all the reconstruction products for astrophysical analysis.


\subsection{Multi-term measurement equation}
Eqn.~(\ref{Eq:combfunc}) can be combined with the measurement process of an imaging interferometer to give the following Measurement equations (written in matrix notation).
\begin{equation}
\sum_{r=0}^{N_r}[\A_r]\I_r = \V
\label{Eq:meqn1}
\end{equation}
where $\I_r$ is an $m\times 1$ list of pixel amplitudes that represents $\{ \I^{sky}_{s,0},  \I^{sky}_{s,p},  \I^{sky}_{s,q} ~\forall s,p,q \}$ as written in Eqn.~(\ref{Eq:combfunc}). Each $[\A_r]$ is an $n\times m$ measurement matrix for the quantity $\I_r$ which produces a list of $n$ visibility measurements ($n = n_{baselines}\times n_{timesteps} \times n_{channels}$). Ignoring direction-dependent instrumental effects, we can write $[\A_r] =[W_r][S][F]$ where $[F]$ is an $m\times m$ Fourier-transform operator, $[S]$ is an $n\times m$ matrix of ones and zeros representing the $uv$-coverage for all baselines over frequency and time, and $[W_r]$ is an $n\times n$ diagonal matrix of weights that describe the basis functions $\I^{shp}_s$, $\dnuno^p$ and $\dtmto^q$.  A set of $N_r$ such systems are added together to form a model of the $n$ measured visibilities.

The measurement equations in block matrix form (for $N_r=3$) are given by 
\begin{equation}
\left[\begin{array}{lll} 
\noalign{\medskip}
   [\A_0] & [\A_1] & [\A_2]  \\
\noalign{\medskip}
   \end{array} \right]
\left[\begin{array}{l}\I_0 \\ \I_1 \\ \I_2\end{array}\right] = \V
\label{eq:meqn}
\end{equation}
All $N_r$ measurement matrices of shape $n\times m$ 
are placed side-by-side to form a larger measurement matrix.
The list of parameters becomes a vertical stack of $N_r$ vectors each of shape
$m\times 1$. The new measurement matrix of shape $n\times m N_r$
operates on
an $m N_r \times 1$ list of parameters to form an $n\times 1$ list of measurements.

A least-squares solution of this system of equations requires the normal equations to be constructed and solved as described below.

\subsection{Multi-term normal equations}
The normal equations (for $N_r=3$) are given by 
\begin{equation}
\left[\begin{array}{lll} 
\noalign{\medskip}
\noalign{\medskip}
   [\Ad_0\W\A_0] & [\Ad_0\W\A_1] & [\Ad_0\W\A_2] \\  
\noalign{\medskip}
   [\Ad_1\W\A_0] & [\Ad_1\W\A_1] & [\Ad_1\W\A_2] \\
\noalign{\medskip}
   [\Ad_2\W\A_0] & [\Ad_2\W\A_1] & [\Ad_2\W\A_2] \\
\noalign{\medskip}
\noalign{\medskip}
   \end{array} \right]
\left[\begin{array}{l}
\noalign{\medskip}
\noalign{\medskip}
\I_0 \\ 
\noalign{\medskip}
\I_1 \\ 
\noalign{\medskip}
\I_2 \\
\noalign{\medskip}
\noalign{\medskip}
\end{array}\right] =
\left[\begin{array}{l}
\noalign{\medskip}
\noalign{\medskip}
[\Ad_0\W]\V \\ 
\noalign{\medskip}
[\Ad_1\W]\V \\ 
\noalign{\medskip}
[\Ad_2\W]\V\\
\noalign{\medskip}
\noalign{\medskip}
\end{array}\right]
\label{eq:neqn} 
\end{equation}
where $[\W]$ is an $n\times n$ diagonal matrix of data weights, usually based on system noise levels.
The Hessian matrix on the LHS consists of $N_r\times N_r$ blocks, each of size $m\times m$.
The list of parameters is a stack of $N_r$ column vectors $\I_r$, and 
the RHS vector is a set of $N_r$ weighted inversions of the data vector.

When $[\A_r] =[W_r][\Sa][F]$, each Hessian block $[\Ad_{r_i}\W\A_{r_j}]$ is a
Toeplitz convolution operator, with each row containing a shifted version of a
point-spread-function\cite{IEEE_CALIM_2009, URV_THESIS} .
Each image vector in the RHS is therefore a sum of convolutions of different image pixel vectors and convolution kernels, and the process of reconstructing $\I_r ~ \forall r$ is equivalent to a joint deconvolution. 

This system of equations is evaluated by first computing the RHS vectors and one point-spread-function per Hessian block. 

The point spread function for Hessian block $r_i,r_j$ is computed as 
$[\Fd\Sd\W_{r_i}\W\W_{r_j}]\vec{1}$. This translates to the 
inverse Fourier transform of the product of the $uv$-coverage, data weights $[\W]$ 
and the spatial-frequency signature of the basis functions labeled by $r_i,r_j$.  

The RHS vector contains a stack of $N_r$ residual images, each 
computed as $[\Fd\Sd\Wd_r\W]\V$.
This translates to weighting the visibilities using data weights as well as weights that
evaluate the $r^{th}$  basis function (a matched-filtering step), 
resampling the visibilities onto a regular grid and then taking an inverse Fourier transform. 
It is important to note that 
basis-function weights for the time and frequency axes must be applied in the 
spatial-frequency domain, prior to resampling the list of visibilities onto a
regular grid. This is to ensure that if multiple visibility values from different
baselines, frequencies and times get mapped onto a single $uv$ grid cell, 
all the contributing visibilities have the correct weights applied to them. 
Weights for the multi-scale basis functions, however, can be applied as a
multiplication on the gridded visibilities.

\paragraph{Differences with the Sault-Wieringa approach :}
The Sault-Wieringa multi-beam algorithms of Refs.~\citenum{MFCLEAN} and \citenum{MBCLEAN2011} constructs RHS vectors by convolving the standard residual image with point-spread-functions constructed for each series term ($I^{rhs}_{r} = I^{rhs}_{0} \star I^{psf}_r$). In the visibility-domain, this corresponds to the multiplication of the gridded visibilities with gridded weights.  If there are any $uv$ grid cells with contributions from measurements from different frequencies, they do not get their correct individual weights.  
The Sault-Wieringa approach works well with interferometers with minimal overlap in 
spatial-frequency coverage (as demonstrated by Ref.~\citenum{MFCLEAN} with the ATCA and Ref.~\citenum{MBCLEAN2011} with e-MERLIN, but 
incurs errors for arrays with dense spatial-frequency coverage such as the EVLA.
For multi-scale multi-frequency imaging, changing the computations to follow those
discussed here and in Ref.~\citenum{MSMFS2011} eliminated this instability and allowed reconstructions
of the source spectrum for extended emission at an accuracy sufficient for
astrophysical analysis. 
A similar argument holds for the SW-based algorithm described in Ref.~\citenum{MBCLEAN2011} when 
solving for time-variability.

\subsection{Solving the Normal equations}
The normal equations shown in Eqn.~(\ref{eq:neqn}) are solved iteratively.

First, the pseudo-inverse of the Hessian is computed by inverting a block-diagonal approximation of the matrix for the time-dependent and frequency-dependent terms  and a diagonal approximation for the multi-scale terms (required for numerical stability\cite{MSMFS2011}).

This pseudo-inverse is applied to all pixels of the RHS images. 
A set of $N_r$ solutions is then picked by finding the location and scale-size corresponding to the flux-components that produces the largest reduction in $\chi^2$ in the current step.
 The residual images are updated by evaluating the LHS for each set of components and subtracting from the RHS.  
Iterations of the above steps are performed until the peak residual of the zero-th order point-source image reaches the level of the first sidelobe of the zero-th order point-spread function. Model visibilities are calculated by evaluating Eqn.~(\ref{eq:meqn}), residual visibilities and RHS residual images are constructed, and the process repeats until residuals are noise-like.
 These steps are  a direct extension of the normalization, peak-finding and update steps of the standard CLEAN\cite{CLEAN} algorithm, and are very similar to the steps of the MS-CLEAN\cite{MSCLEAN} , SW-CLEAN\cite{MFCLEAN} and SW-Multi-Beam\cite{MBCLEAN2011} algorithms. 

\paragraph{Software Implementation : }
The MS-MF-TV-CLEAN algorithm described here was implemented as an extension of the existing MS-MFS algorithm using the CASA software libraries. 
The following acronyms will be used to distinguish between various subsets of this implementation used for the tests in Sec.~\ref{sec:examples}. 
\begin{enumerate}
\item MS-CLEAN : The model contains only multi-scale terms. The resulting algorithm differs from that in Ref.~\citenum{MSCLEAN} in that it does not require a scale-dependent bias-factor.
\item MS-MF-CLEAN : The model contains frequency-dependent terms with multi-scale components\cite{MSMFS2011} .
\item MS-TV-CLEAN : The model contains time-variable terms with multi-scale components. The time-variability part of the algorithm differs from that in Ref.~\citenum{MBCLEAN2011} in its choice of basis functions and how it computes residual images.
\item MS-MF-TV-CLEAN : The model contains time and frequency-variable terms with multi-scale components.  The MF-TV part of this method differs from that in Ref.~\citenum{MBCLEAN2011} in the way the time and frequency basis functions are combined and uses fewer total terms in the series expansion of the model.
\end{enumerate}


\section{Imaging 3D structure in the solar corona}
\label{sec:examples}

Gyro-synchrotron radio emission from the solar corona within the frequency range of 1 GHz to 20 GHz traces magnetic-field strengths at difference heights above the surface of the sun\cite{CorMagFASR} . 
Details about this 3D magnetic-field structure and how it varies with time are essential to the understanding of the physics driving solar flares and other coronal activity. Radio interferometric observations at cm wavelengths provide crucial information for theoretical models, and 
several recent efforts\cite{Fleishman2009, Fleishman2010Concept, Fleishman_modeling2011} have made progress on forward-modeling using reconstructed multi-frequency image cubes. 

The image-reconstruction itself, however, has been limited to traditional methods of making images at multiple separate frequencies, using narrow-band data, and using data accumulated over time-durations shorter than the typical variation timescale.   With such narrow-band, snapshot interferometric data, the non-linear reconstruction process has to work with a limited number of samples that is often insufficient for an un-ambiguous reconstruction of the complicated extended emission present in active coronal regions. Further, when data from different frequencies are treated separately, the intrinsic frequency-dependent angular resolution of the imaging instrument comes into play, and all further analysis of the resulting image cubes must be done at the angular resolution offered by the lowest frequency. This can be improved using a wideband imaging algorithm\cite{MSMFS2011} that reconstructs frequency-dependent structure, but it is still restricted to only snapshot $uv$-coverages.

An algorithm such as that described in this paper has the potential of reconstructing spatially-extended structure that varies smoothly in shape with both time and frequency within the parameters of a synthesis observation. Such an algorithm would be able to take advantage of the increased imaging-fidelity that comes from the better spatial-frequency coverage generated by Earth-rotation-synthesis and multi-frequency-synthesis, and not be overly restricted by the intrinsic angular resolution at different frequencies.

The following sections describe two examples in which radio interferometric observations of 3D structures in the solar corona are simulated, and then reconstructed using both traditional and new methods. The goal of these tests is to assess if the image-models described in Sec.~\ref{sec:algorithms} suffice to reconstruct time and frequency-dependent extended-structure at an accuracy sufficient for further analysis, at an angular resolution better than that given by the lowest measured frequency.

\subsection{Frequency-dependent structure - a solar flare loop}
\label{sec:loops}

\begin{figure}
   \begin{center}
   \begin{tabular}{cc}
   \includegraphics[height=4.5cm]{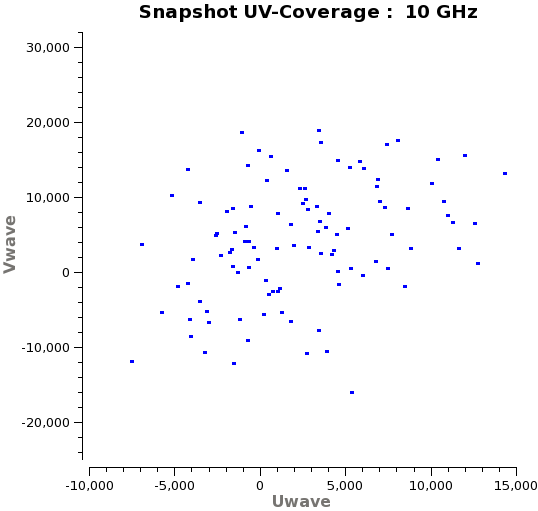}
   \includegraphics[height=4.5cm]{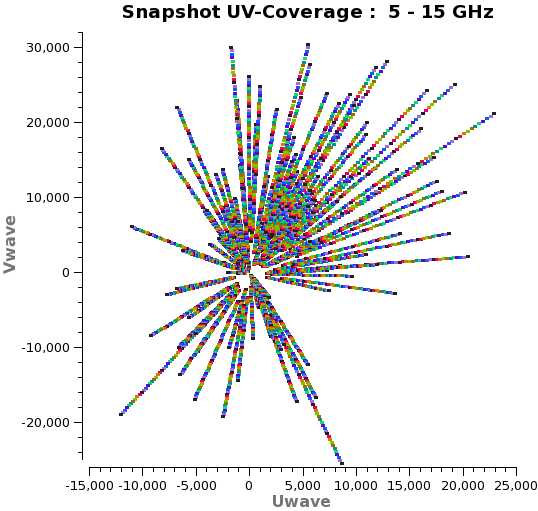}
   \end{tabular}
   \end{center}
   \caption[Spatial-frequency coverage plots - 1] 
   { \label{fig:fasr_15ant_uvcov} The snapshot narrow-band spatial-frequency coverage at 10 GHz (LEFT) is compared with the broad-band coverage between 5 GHz and 15 GHz, sampled at intervals of 50 MHz (RIGHT). This coverage corresponds to a 15 element array, following a randomized log-spiral pattern with upto 400m baselines. This narrow-band coverage is considered sparse for structures seen for solar flare loops, and any algorithm that uses the broad-band coverage will have an advantage.  }
   \end{figure}

Interferometric data for a solar-loop model were simulated for a snapshot observation with a 15-element array with broad-band receivers covering the range of 5 GHz to 15 GHz. The antennas follow a randomized log-spiral pattern with up to 400m baselines and represents the inner region of the proposed FASR telescope\cite{FASRImaging2003} . Fig.~\ref{fig:fasr_15ant_uvcov} compares the spatial-frequency coverage of the interferometer at 10 GHz with the wideband spatial-frequency coverage, both for a single snapshot observation.   The intrinsic angular resolution of this instrument is controlled by the largest sampled spatial-frequency and varies from 40 arcsec at 5 GHz to 10 arcsec at 15 GHz. 

\begin{figure}
   \begin{center}
   \begin{tabular}{ccc}
   \includegraphics[height=2.7cm]{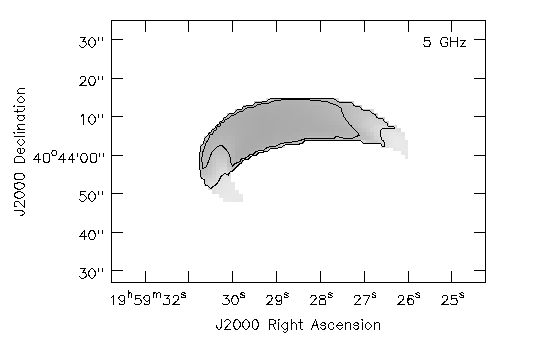}
   \includegraphics[height=2.7cm]{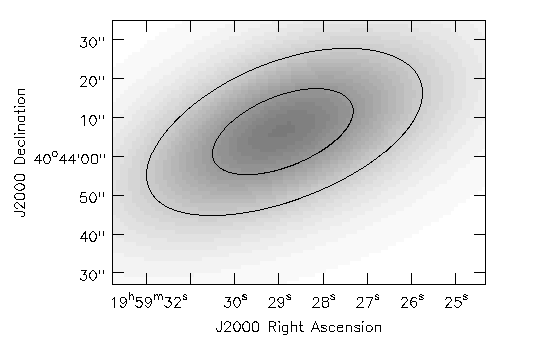}
   \includegraphics[height=2.7cm]{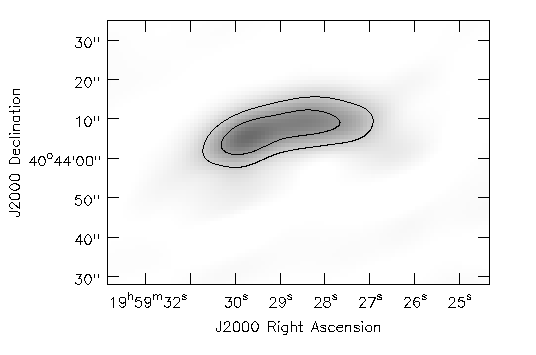} \\
   \includegraphics[height=2.7cm]{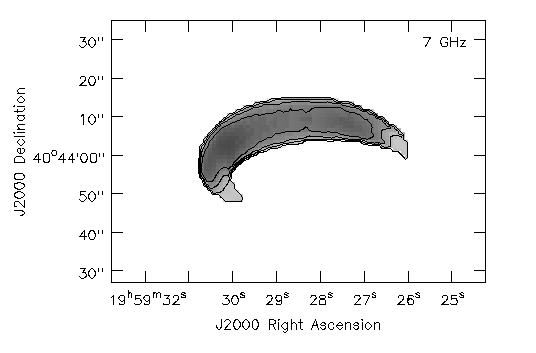}
   \includegraphics[height=2.7cm]{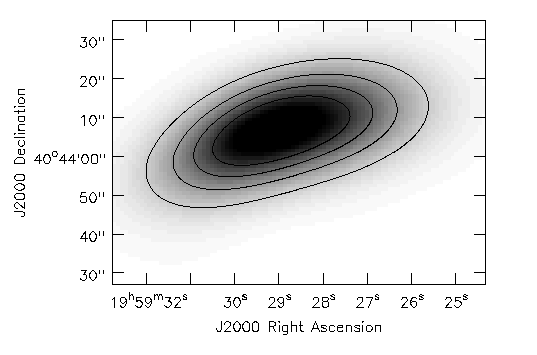}
   \includegraphics[height=2.7cm]{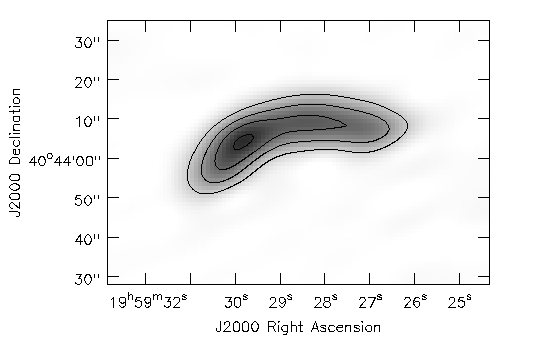} \\
   \includegraphics[height=2.7cm]{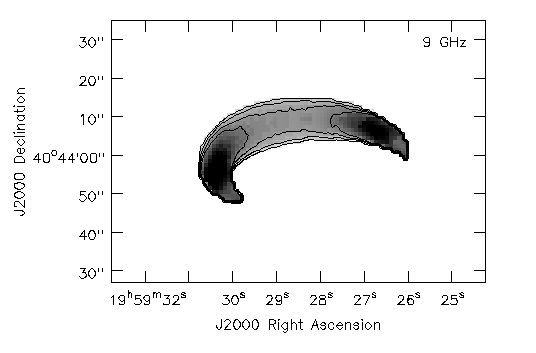}
   \includegraphics[height=2.7cm]{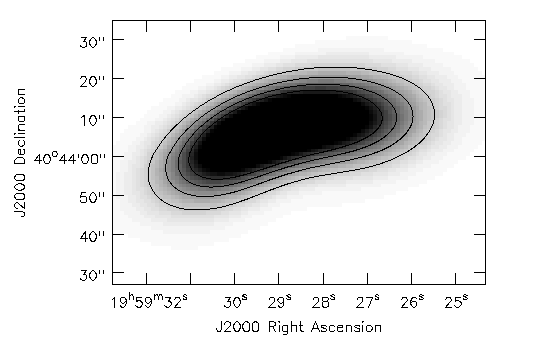}
   \includegraphics[height=2.7cm]{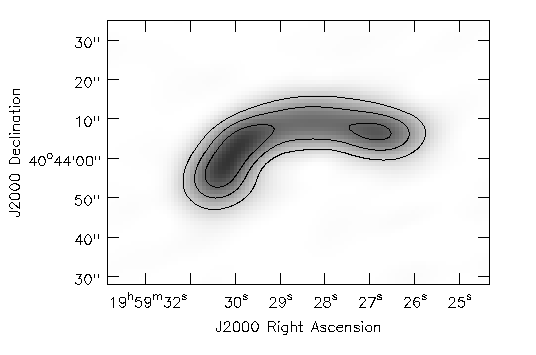} \\
   \includegraphics[height=2.7cm]{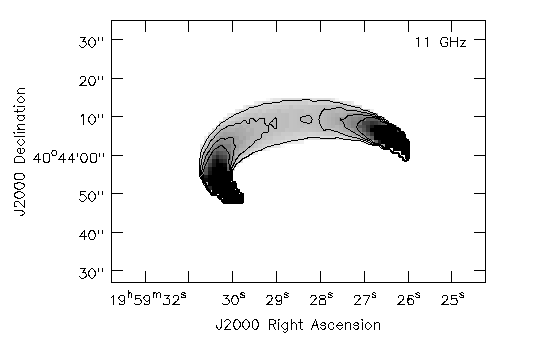}
   \includegraphics[height=2.7cm]{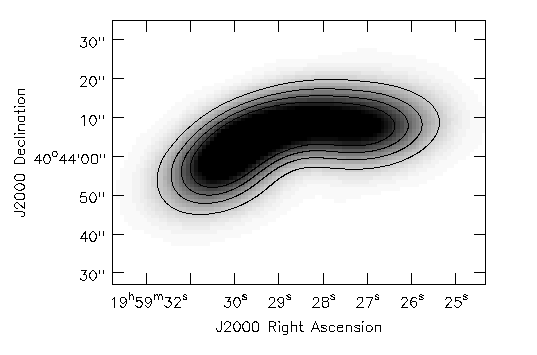}
   \includegraphics[height=2.7cm]{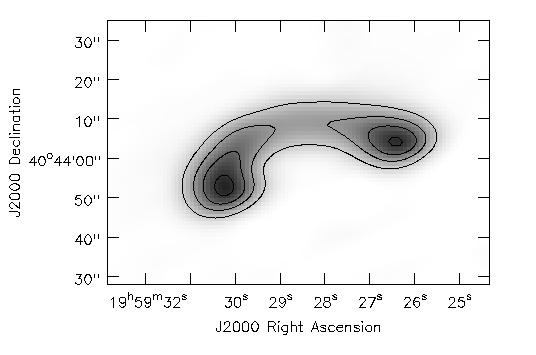} \\
   \includegraphics[height=2.7cm]{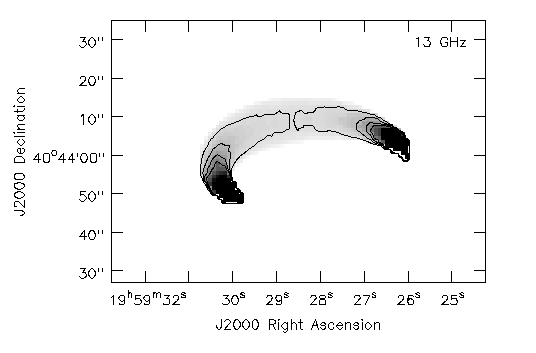}
   \includegraphics[height=2.7cm]{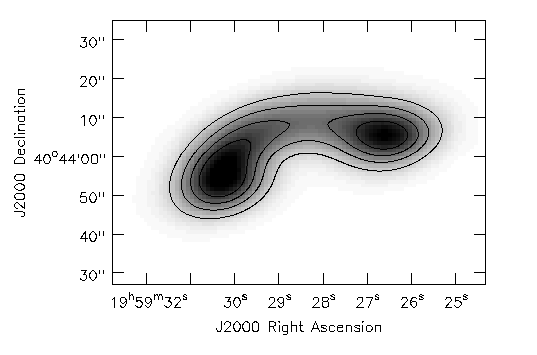}
   \includegraphics[height=2.7cm]{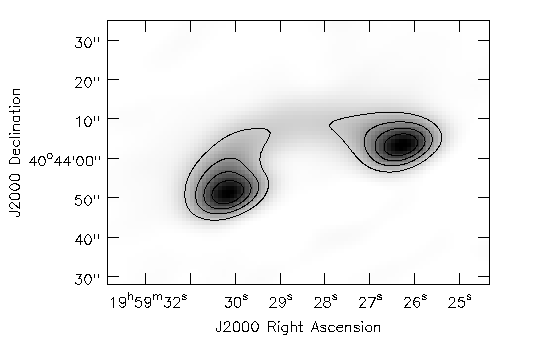} \\
   \includegraphics[height=2.7cm]{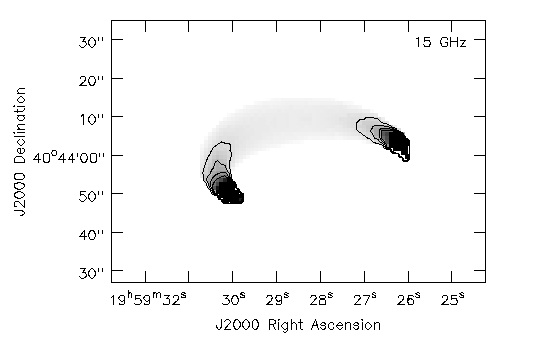}
   \includegraphics[height=2.7cm]{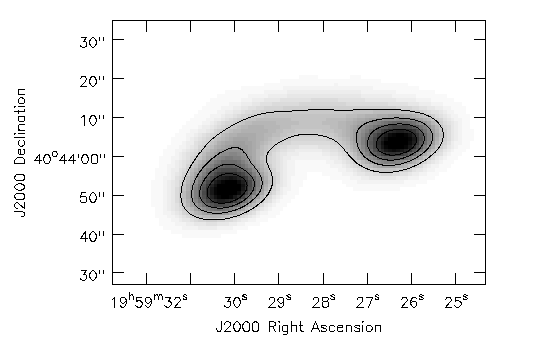}
   \includegraphics[height=2.7cm]{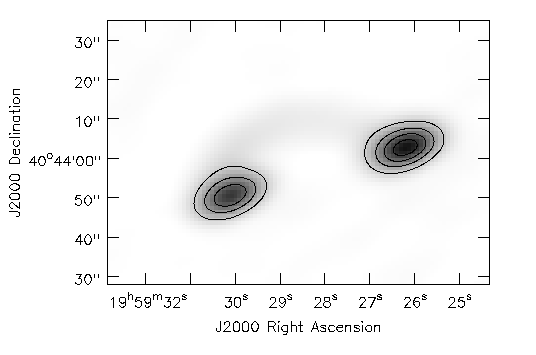} 
   \end{tabular}
   \end{center}
   \caption[Solar Loop Model - comparison of true and reconstructed images] 
   { \label{fig:loops} The frequency-dependent structure of a solar flare loop at 5 GHz, 7 GHz, 9 GHz, 11 GHz and 13 GHz (ROWS : Top to Bottom) are shown for the simulated model (LEFT column), reconstructions performed one channel at a time (MIDDLE column), and the wide-band reconstruction using the MS-MF-CLEAN algorithm (RIGHT column).  Moving from lower to higher frequencies (top to bottom) this structure traces the top of the loop and moves deeper in the solar corona until the 'feet' of the loop.  The angular resolutions of the images in the MIDDLE column range from 52''x28'' at 5 GHz to 14''x7'' at 15 GHz, resulting in a plausible reconstruction only at the higher frequencies. For the RIGHT column, the angular resolution is 13''x6'' at all frequencies shown, and the structure is recovered at the same fidelity at all of these frequencies. These reconstructions used the snapshot uv-coverage shown in Fig.~\ref{fig:fasr_15ant_uvcov} and demonstrate the advantage of using the broad-band uv-coverage along with an algorithm that accounts for the changing structure as a function of frequency.}
   \end{figure}

Figure \ref{fig:loops} compares imaging results using traditional and new methods, and compares them with the true images used in the simulation. 
 
The left column shows the true model of a solar-flare loop at 6 different frequencies. This model was constructed such that frequencies around 6 GHz trace structure near the top of the solar-flare loop, and higher frequencies trace structure deeper in the solar corona, down to the 'feet' of the solar-flare loop. 

The middle column shows the reconstruction using the MS-CLEAN algorithm separately on each frequency-channel. This is the traditional form of analysis of such wide-band interferometric data, and the image fidelity is clearly limited by the varying angular resolution of the instrument as a function of frequency.  

The column on the right shows reconstructions built from a multi-scale multi-frequency model whose coefficients were fitted using the MS-MF-CLEAN algorithm on the data combined over all frequencies. With a 3-term Taylor-polynomial, this reconstruction was able to recover the 3D structure at an angular resolution of 10 arcsec.  For the top part of the loop, the algorithm chose flux components whose amplitudes decreased with increasing frequency, and for the feet, it chose components whose amplitudes were low at lower frequencies and increased with frequency.

\subsection{Time and frequency variable structure - an expanding active coronal region}
\label{sec:sunspot}

A model of an active coronal region was generated
such that it shows compact structure in deeper layers (probed by higher frequencies), with the upper layers containing an extended plume.  As a function of time, these upper layers expand spatially, but the lower layers remain relatively unchanged.
Data were simulated for the EVLA between 5 GHz and 15 GHz, to test whether such an experiment would be feasible with an existing telescope. A set of 8 snapshots were simulated, spread across 4 hours. 

\begin{figure}
   \begin{center}
   \begin{tabular}{ccc}
   \includegraphics[height=3.5cm]{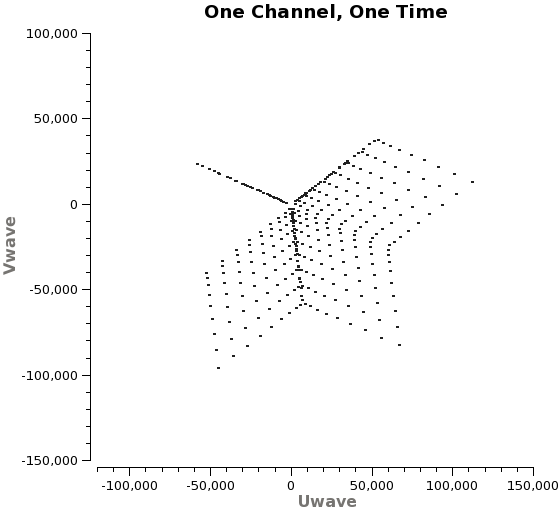}
   \includegraphics[height=3.5cm]{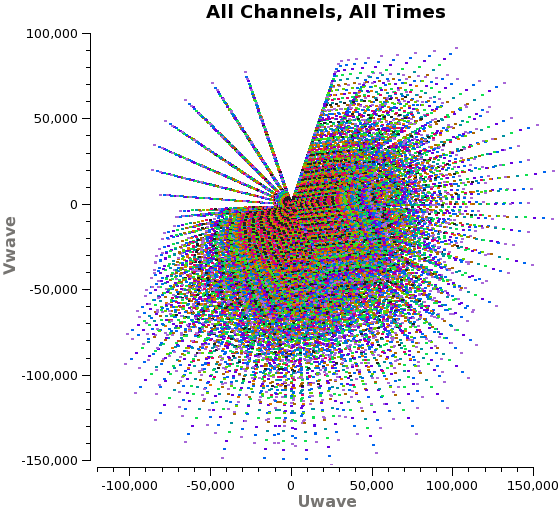}\\

   \includegraphics[height=3.8cm]{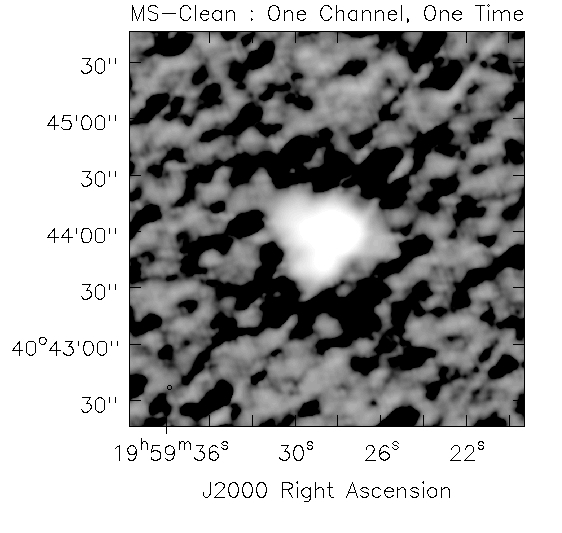}
   \includegraphics[height=3.8cm]{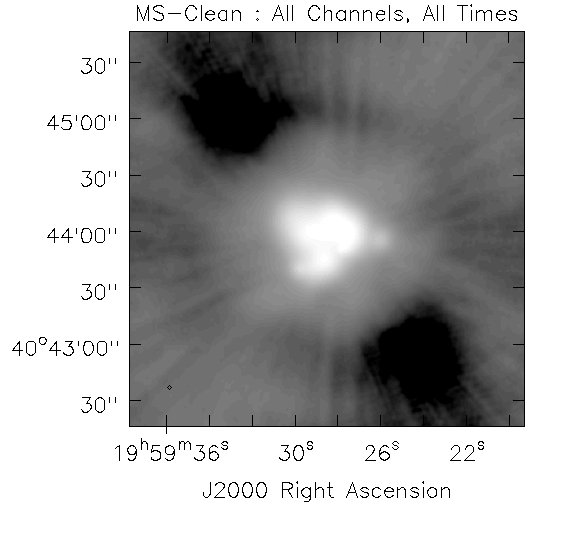}\\

   \includegraphics[height=3.5cm]{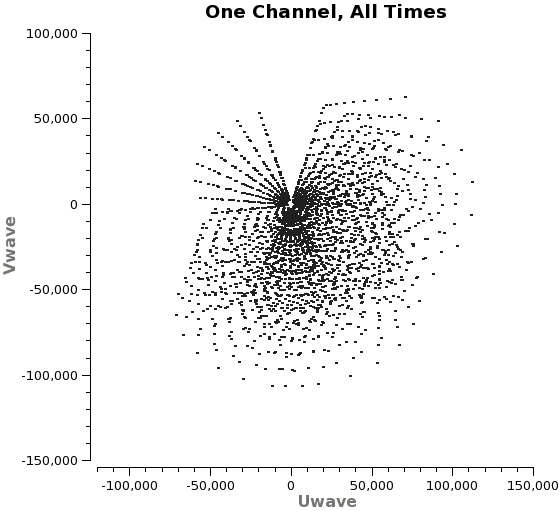}
   \includegraphics[height=3.5cm]{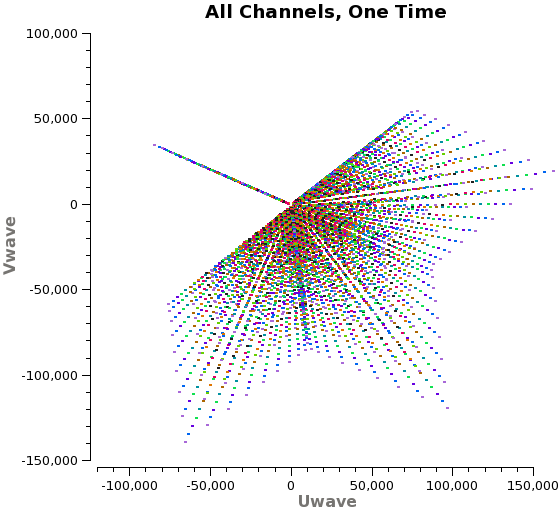}
   \includegraphics[height=3.5cm]{Figures/plot_evla_uvcov_allchan_alltime}\\

   \includegraphics[height=3.8cm]{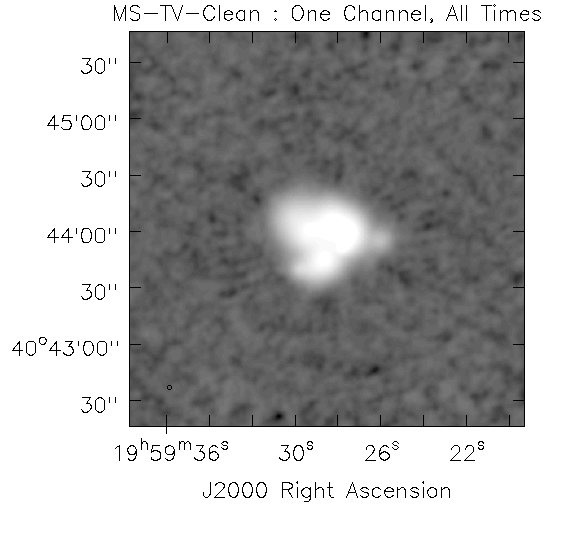}
   \includegraphics[height=3.8cm]{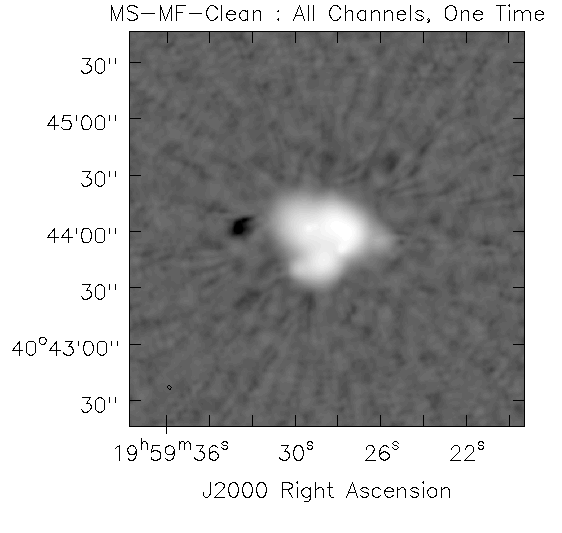}
   \includegraphics[height=3.8cm]{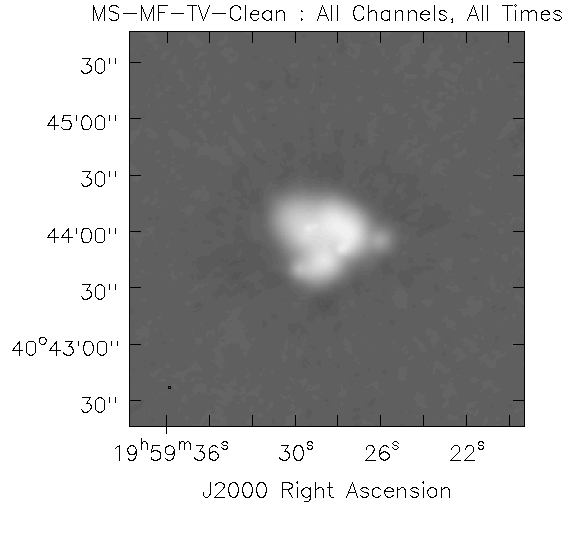}
   \end{tabular}
   \end{center}
   \caption[example] 
   { \label{fig:evla_sunspot_1} This figure compares the results of five reconstructions using different subsets of the data and different algorithms. All images are displayed with the same gray-scale/range. Two plots are shown for each example; the uv-coverage and the reconstructed average intensity (shown below the uv-coverage plots).  The first two examples (top two rows) show results of the MS-CLEAN algorithm, ignoring time and frequency variable structure. The first example (top left) shows the results with narrow-band snapshot uv-coverage, and represents the imaging fidelity achieved via the traditional method of handling time-frequency-variable structure by simply partitioning the data and imaging each frequency and timestep independently.  The second example (top right) shows the results with all the data, taking advantage of Earth-rotation and multi-frequency synthesis but ignoring all variations in time and frequency.   The next three examples (bottom row) show the results of various Multi-Term algorithms. The first example (bottom left) shows an MS-TV-CLEAN reconstruction on one-channel of data while fitting for the time-variability. The second example (bottom middle) shows an MS-MF-CLEAN reconstruction on a multi-channel snapshot, where the frequency-structure is accounted-for. These two examples show some improvement over pure MS-CLEAN, but still show residuals at a level higher than what the dataset allows.  The last example (bottom right) shows the MS-MF-TV-CLEAN reconstruction where data from all channels and timesteps were used and the algorithm fitted simultaneously for time and frequency variations in structure. The noise-level achieved is lower than other examples, but still two orders of magnitude higher than the theoretical estimate. Table~\ref{tab:sunspot_rms} lists off-source RMS noise levels for all 5 examples.\\
   }
   \end{figure}

\begin{figure}
   \begin{center}
   \begin{tabular}{ccp{0.5cm}cc}
   \includegraphics[height=3.3cm]{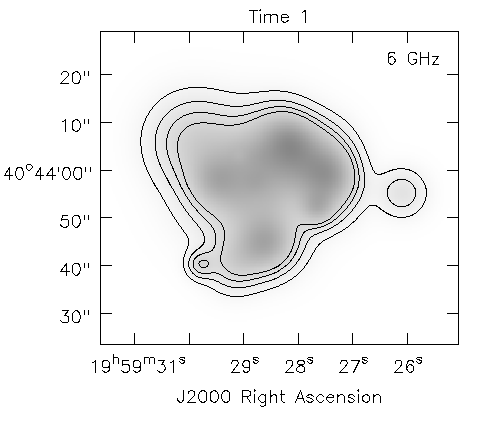} &
   \includegraphics[height=3.3cm]{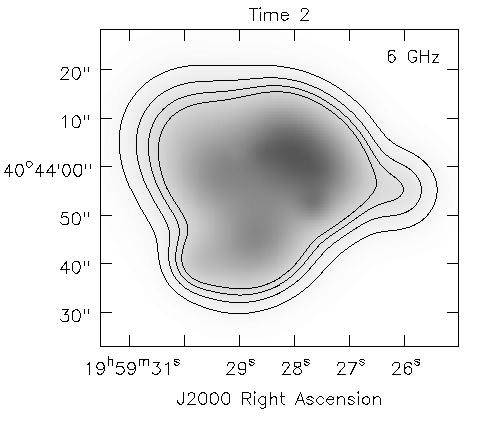} &  &
   \includegraphics[height=3.3cm]{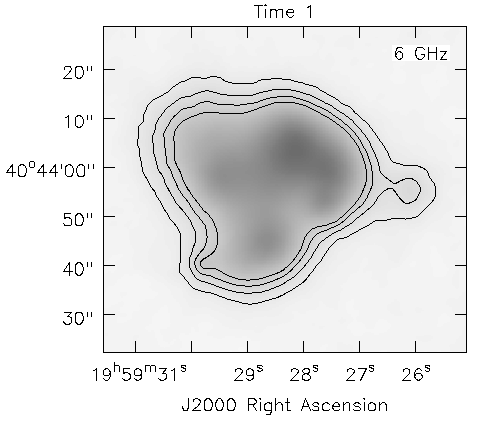} &
   \includegraphics[height=3.3cm]{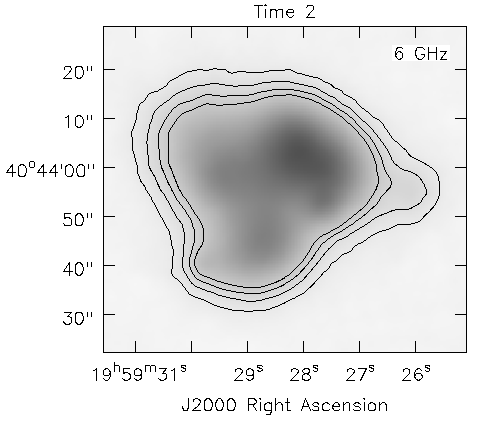} \\
   \includegraphics[height=3.3cm]{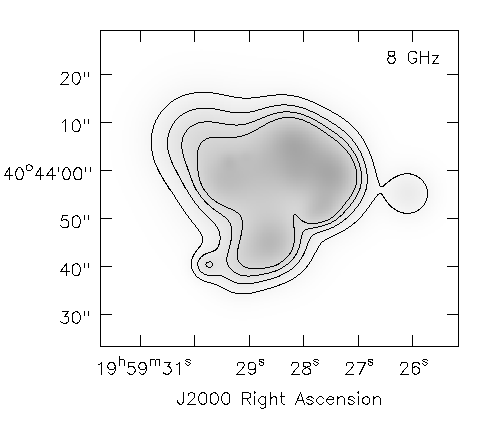} &
   \includegraphics[height=3.3cm]{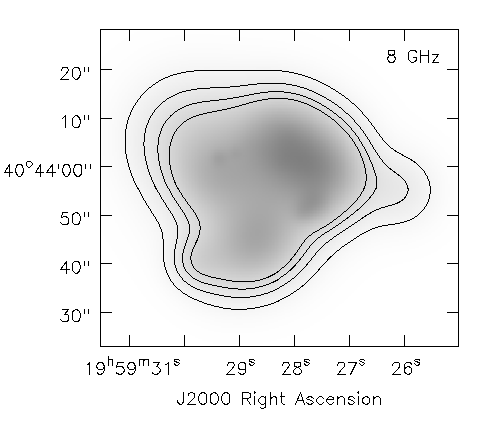} &  &
   \includegraphics[height=3.3cm]{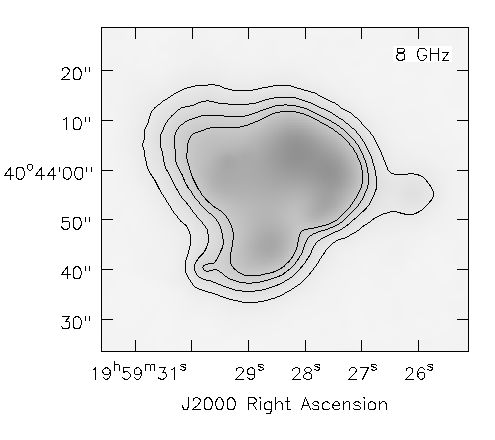} &
   \includegraphics[height=3.3cm]{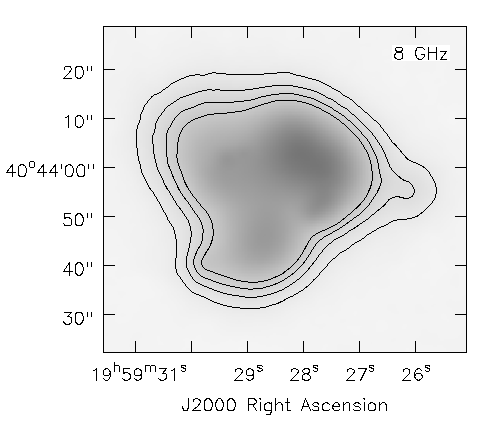} \\
   \includegraphics[height=3.3cm]{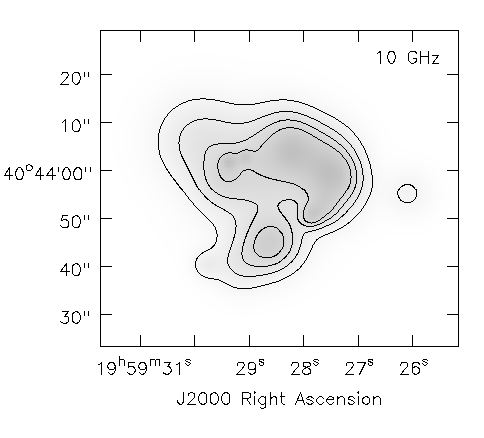} &
   \includegraphics[height=3.3cm]{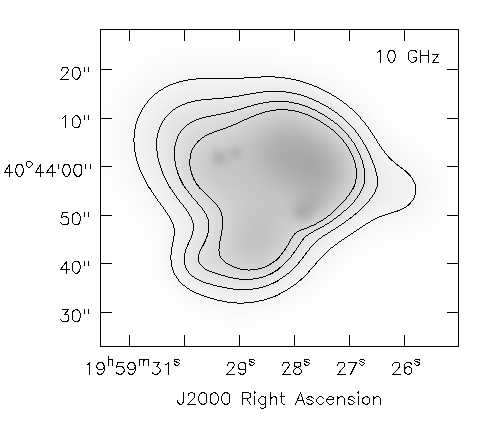} &  &
   \includegraphics[height=3.3cm]{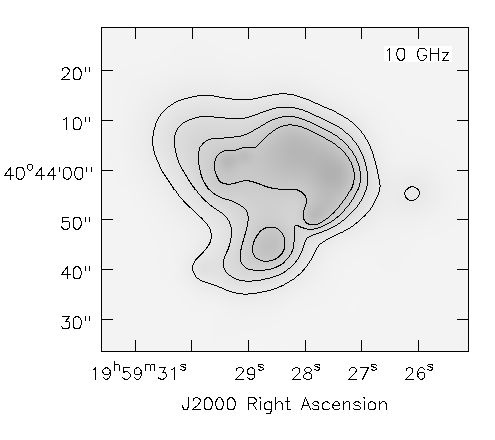} &
   \includegraphics[height=3.3cm]{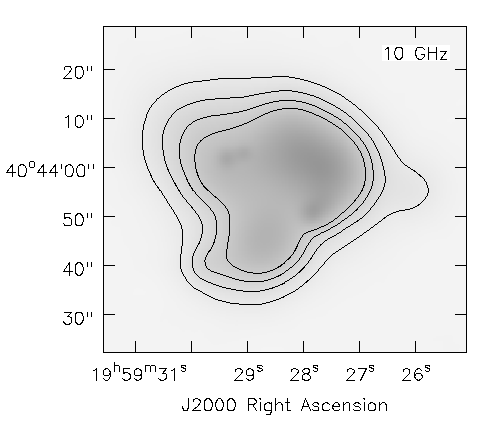} \\
   \includegraphics[height=3.3cm]{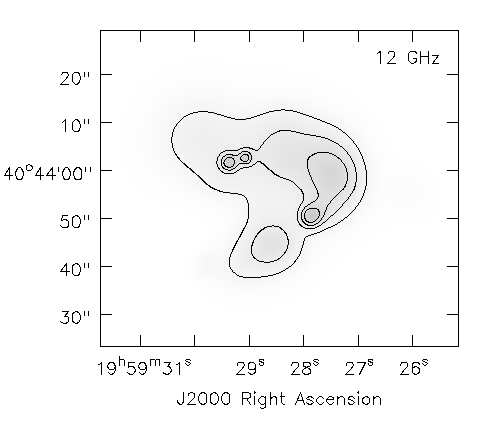} &
   \includegraphics[height=3.3cm]{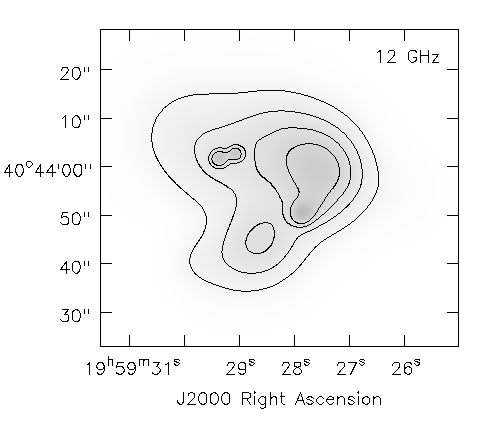} &  &
   \includegraphics[height=3.3cm]{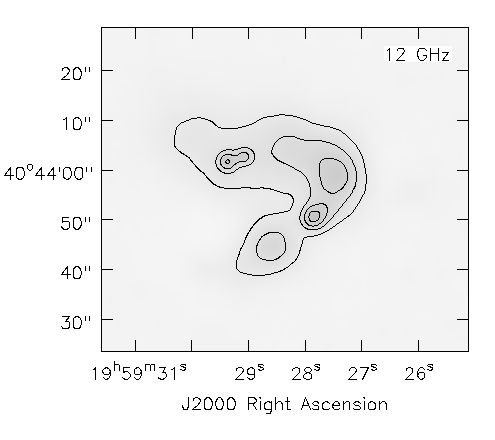} &
   \includegraphics[height=3.3cm]{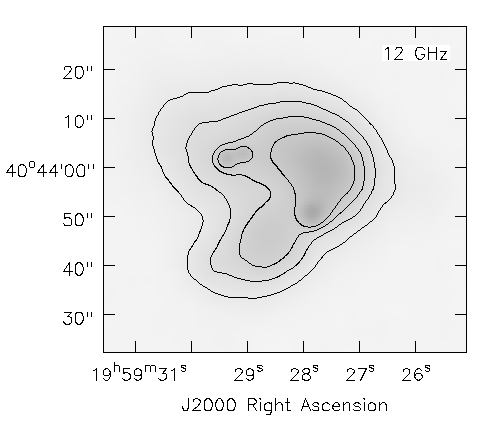} \\
   \includegraphics[height=3.3cm]{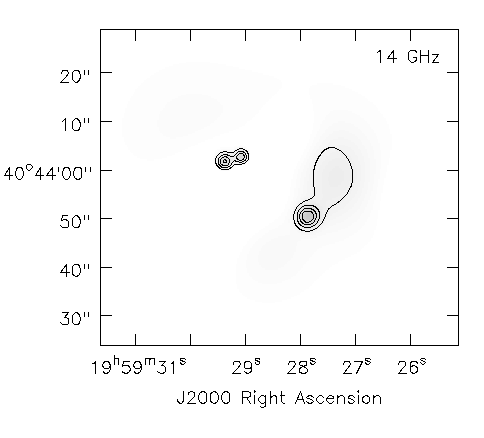} &
   \includegraphics[height=3.3cm]{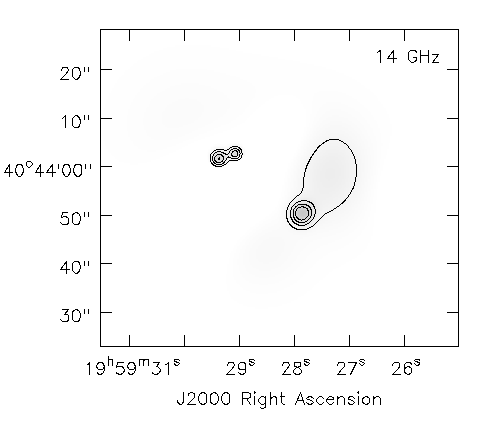} &  &
   \includegraphics[height=3.3cm]{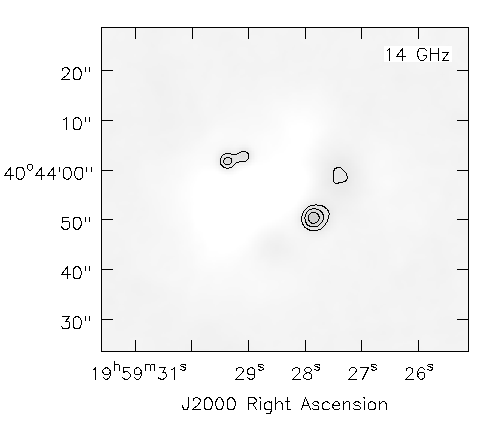} &
   \includegraphics[height=3.3cm]{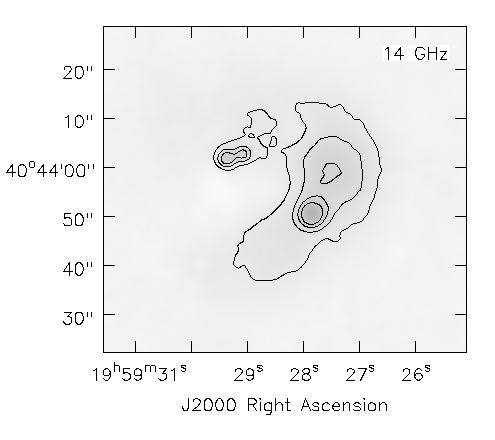} \\
   \end{tabular}
   \end{center}
   \caption[example] 
   { \label{fig:evla_sunspot_2} This figure shows the multi-frequency and time-variable structure of the simulated radio emission from a sunspot, and compares the true structure (COLUMNS 1,2) with that reconstructed via the MS-MF-TV-CLEAN algorithm (COLUMNS 3,4). These results are from the same run represented in last pair of plots in Fig.~\ref{fig:evla_sunspot_2}, and used data from 20 channels and 8 timesteps.  In each pair of columns, frequency ranges from 6 GHz (TOP) to 14 GHz (BOTTOM), and the time difference between each pair of columns is 4 hours. This structure represents the 3D structure of a sunspot whose upper layers are expanding spatially as time progresses.  The reconstructions (COLUMNS 3,4) show that most of the structure has been reconstructed. The sky model solved-for comprised of a linear combination of multi-scale flux-components, each of whose amplitudes had an average intensity term, a gradient along frequency, and a gradient along time.  This simple model resulted in good reconstructions in most of the frequency-range, but showed errors in modeling extended emission at frequencies lower than 6 GHz and higher than 12 GHz.}
   \end{figure}

Figure \ref{fig:evla_sunspot_1} shows $uv$-coverages of various subsets of the data, as well as the average-intensity images produced by applying different algorithms to it. 
Table \ref{tab:sunspot_rms} lists the off-source RMS levels with the different methods, along with the number of data-points and the algorithm used in the reconstruction.
The true model used in the simulations is shown in the first two columns of Fig.~\ref{fig:evla_sunspot_2}.

The results can be summarized as follows. 
The traditional method of imaging each channel and snapshot separately is clearly limited in its dynamic-range and fidelity, even if MS-CLEAN is applied. Also, if MS-CLEAN is applied to the entire dataset without accounting for time and frequency variations, artifacts still dominate the residuals and affect the on-source reconstructionas well.  Intermediate methods of solving for time-variant spatial structure one channel at a time (MS-TV-CLEAN),  or solving for frequency-dependent structure one timestep at a time (MS-MF-CLEAN), yield better results, but the limited $uv$-coverage still results in imaging artifacts.  Finally, a combined reconstruction using MS-MF-TV-CLEAN produces the lowest residuals, although even these numbers are more than an order-of-magnitude larger than theoretical point-source sensitivity estimate for this simulated dataset. The dynamic range of this image is a few hundred to one thousand, which is likely to be sufficient for strong radio emission from the sun. 

   \begin{table}
   \begin{center}
     \begin{tabular}{|l|l|l|l|l|}
       \hline
   Aperture Synthesis                 &      N-channels & N-timesteps & Algorithm & Off-source RMS \\
\hline
 Single-frequency + Snapshot            &    1               &         1        & MS-CLEAN & 3.5 mJy \\       
 Multi-frequency + Earth-rotation   &      20               &       8        & MS-CLEAN & 1.4 mJy \\       
 Single-frequency + Earth-rotation           &        1               &         8        & MS-TV-CLEAN & 0.43 mJy \\       
 Multi-frequency  + Snapshot &    20               &       1        & MS-MF-CLEAN & 0.3 mJy \\       
 Multi-frequency + Earth-rotation &       20               &        8        & MS-MF-TV-CLEAN & 0.075 mJy \\       
Multi-frequency + Earth-rotation &        20               &        8        & Theoretical RMS & 0.0004 mJy  \\
\hline   
      \end{tabular}
      \end{center}
        \caption[RMS comparison]{\label{tab:sunspot_rms} This table lists off-source RMS values in the reconstructed average intensity images for the five scenarios discussed in Sec.~\ref{sec:sunspot} and illustrated in Fig.~\ref{fig:evla_sunspot_1}. Algorithms that model time and/or frequency variant structure are able to take advantage of the extra spatial-frequency coverage offered by the combining the data, and reduce the level at which imaging-artifacts occured.}
        \end{table}


Figure \ref{fig:evla_sunspot_2} compares the true source structure with that reconstructed via MS-MF-TV-CLEAN with 2-terms along the frequency axis and 2-terms along the time axis, and 5 different spatial scales. The images in the 3rd and 4th columns show that most of the time-variable 3D structure has been recovered.  This is a promising result, but the reconstructions  at the highest frequencies show some spurious extended emission. This points to either the inadequacy of a 2-term model for the source 'spectrum', or the inherent ambiguity\cite{MSMFS2011} in the spectral signature at spatial scales that are not sampled at all observed frequencies.

\section{Discussion}

Image-reconstruction algorithms for radio interferometry that model time and frequency structure can be used not-only to improve the fidelity of the average image, but to also reconstruct the frequency-dependent and time-variable structure at an accuracy sufficient for astrophysical interpretation.   Smooth frequency-dependent structure can be reconstructed across frequency ranges wider than previously demonstrated, achieving higher angular-resolution than that given by the lowest frequency in the measured band.  Reconstructing time-variability is also possible, but since angular-resolution is not of concern here, this feature may be useful only in the context of improving signal-to-noise and image-fidelity by using all measurements together.

The example in Sec.~\ref{sec:sunspot} is a proof-of-concept demonstration that such imaging of frequency-dependent and time-variable structures in coronal active regions may be possible with current telescopes such as the EVLA.  Future telescopes such as FASR will have more than enough spatial-frequency coverage for an unambiguous reconstruction at each timestep and channel, even after tapering data from all frequencies to a common low angular resolution.

From a practical standpoint, the MS-MF-TV-CLEAN implementation described in this paper improves upon the multi-beam algorithm in Ref.~\citenum{MBCLEAN2011} in that it does not suffer from instabilities incurred when there is significant overlap between sampled spatial-frequencies. It suggests the choice of $N_p + N_q - 1$ basis functions instead of the $N_p \times N_q$, and 
also folds-in multi-scale support in a way that allows a wide choice of spatial basis functions whose amplitudes are given by time and frequency polynomials. 

There is always a trade-off between a basis set that optimally represents the structure being reconstructed and requires a minimal number of components to model it, and a basis set whose members the instrument can optimally distinguish between (an orthogonal basis for the instrument) but may require many more components to model the target structure.  For the situations explored in this paper, very simple polynomial basis functions produce usable reconstructions, although some care must be taken to ensure that the instrument can distinguish between them. This information is encoded in the diagonals of each block of the Hessian matrix, and scale-sizes and polynomial orders must be chosen such that its condition number is not too large.
Fine structures in the dynamic spectra can be reconstructed only with higher-order terms which are increasingly harder for the instrument to distinguish from one another, and as demonstrated in Ref.~\citenum{MBCLEAN2011} may require an explicit orthogonalization step. However, this orthogonalization must be restricted to only the time and freqency basis functions because the multi-scale basis functions are usually chosen based on structures that optimally describe the structure being imaged, and a basis function orthogonalization may result in a much larger number of components being required to fit the data.  

One problem with this approach is the current inability to derive accurate error-estimates on the fitted parameters, because it depends not-only on the basis functions and the instrument (as encoded in the Hessian and covariance matrices), but also on the appropriateness of the chosen model for the structure being reconstructed. For instance with MS-MF-CLEAN\cite{MSMFS2011} , errors in the spectral-index data product are as high as a few hundred percent when delta-functions are used to model extended emission.  There is also an inherent ambiguity in the spectral structure at the largest scales, because of the frequency-scaling of the instruments $uv$-coverage, and in this case, the errors are dominated by the inability of the measurements to constrain the model and may require {\it a-priori} information\cite{MSMFS2011, FASRImaging2003} . To be practically usable, better measures of the reconstruction quality are required.

The ideas presented in this paper are one step closer to the direct modeling goals presented in Ref.~\citenum{Fleishman2010Concept} for the reconstruction of 3D magnetic-field structures in the solar corona . 
The next step for such algorithms ought to be to step away from pattern-modeling, and fold physical models directly into the reconstruction process. Refs.~\citenum{Fleishman2009} and \citenum{Fleishman_modeling2011} ,
among others, demonstrate that physical models fitted to single-channel images from an idealized interferometer are capable of recovering magnetic-field structures and other information required to undestand the physical processes. An algorithm that combines these ideas and reconstructs the physical model directly from visibilities would eliminate the intermediate non-linear step of imaging which has its own uncertainties.

\subsection{Acknowledgments} 
The author would like to thank Tim Bastian (NRAO) for providing the loop model for 
the tests described in Sec.~\ref{sec:loops}, and the NRAO-CASA group for access to 
their libraries for the software implementation of the MS-MF-TV-CLEAN algorithm.

\bibliography{tfimaging}   

\begin{thebibliography}{10}

\bibitem{NRAO_LECTURES}
G.~B. {Taylor}, C.~L. {Carilli}, and R.~A. {Perley}, eds., {\em Astron. Soc.
  Pac. Conf. Ser. 180: Synthesis Imaging in Radio Astronomy II}, 1999.

\bibitem{MFCLEAN_CCW}
J.~E. {Conway}, T.~J. {Cornwell}, and P.~N. {Wilkinson}, ``{Multi-Frequency
  Synthesis - a New Technique in Radio Interferometric Imaging},'' {\em
  mnras}~{\bf 246}, p.~490, Oct. 1990.

\bibitem{MFCLEAN}
R.~J. {Sault} and M.~H. {Wieringa}, ``{Multi-frequency synthesis techniques in
  radio interferometric imaging.},'' {\em Astron. \& Astrophys. Suppl.
  Ser.}~{\bf 108}, pp.~585--594, Dec. 1994.

\bibitem{MSMFS2011}
U.~{Rau} and T.~J. {Cornwell}, ``{A multi-scale multi-frequency deconvolution
  algorithm for synthesis imaging in radio interferometry},'' {\em Astron. \&
  Astrophys.}~{\bf 532}, p.~A71, Aug. 2011.

\bibitem{MBCLEAN2011}
I.~M. {Stewart}, D.~M. {Fenech}, and T.~W.~B. {Muxlow}, ``{A multiple-beam
  CLEAN for imaging intra-day variable radio sources},'' {\em aap}~{\bf 535},
  p.~A81, Nov. 2011.

\bibitem{CLARK_CLEAN}
B.~G. {Clark}, ``An efficient implementation of the algorithm 'clean','' {\em
  Astron. \& Astrophys.}~{\bf 89}, p.~377, Sept. 1980.

\bibitem{CLEAN}
J.~A. H\"ogbom, ``Aperture synthesis with a non-regular distribution of
  interferometer baselines,'' {\em Astron. \& Astrophys. Suppl. Ser.}~{\bf 15},
  pp.~417--426, 1974.

\bibitem{MEM}
T.~J. {Cornwell} and K.~J. {Evans}, ``A simple maximum entropy deconvolution
  algorithm,'' {\em Astron. \& Astrophys.}~{\bf 143}, pp.~77--83, 1985.

\bibitem{MEM_RN}
R.~{Narayan} and R.~{Nityananda}, ``{Maximum entropy image restoration in
  astronomy},'' {\em Ann. Rev. Astron. Astrophys.}~{\bf 24}, pp.~127--170,
  1986.

\bibitem{MSCLEAN}
T.~J. {Cornwell}, ``{Multi-Scale CLEAN deconvolution of radio synthesis
  images},'' {\em IEEE Journal of Selected Topics in Sig. Proc.}~{\bf 2},
  pp.~793--801, Oct 2008.

\bibitem{Asp_Clean}
S.~{Bhatnagar} and T.~J. {Cornwell}, ``{Scale sensitive deconvolution of
  interferometric images. I. Adaptive Scale Pixel (Asp) decomposition},'' {\em
  Astron. \& Astrophys.}~{\bf 426}, pp.~747--754, Nov. 2004.

\bibitem{SPARSERI2010}
S.~{Wenger}, M.~{Magnor}, Y.~{Pihlstr{\"o}m}, S.~{Bhatnagar}, and U.~{Rau},
  ``{SparseRI: A Compressed Sensing Framework for Aperture Synthesis Imaging in
  Radio Astronomy},'' {\em pasp}~{\bf 122}, pp.~1367--1374, Nov. 2010.

\bibitem{LICS2011}
F.~{Li}, T.~J. {Cornwell}, and F.~{de Hoog}, ``{The application of compressive
  sampling to radio astronomy. I. Deconvolution},'' {\em aap}~{\bf 528},
  p.~A31, Apr. 2011.

\bibitem{SARA2012}
R.~E. {Carrillo}, J.~D. {McEwen}, and Y.~{Wiaux}, ``{Sparsity Averaging
  Reweighted Analysis (SARA): a novel algorithm for radio-interferometric
  imaging},'' {\em ArXiv e-prints} , May 2012.

\bibitem{IEEE_CALIM_2009}
U.~{Rau}, S.~{Bhatnagar}, M.~A. {Voronkov}, and T.~J. {Cornwell}, ``{Advances
  in Calibration and Imaging Techniques in Radio Interferometry},'' {\em IEEE
  Proceedings}~{\bf 97}, pp.~1472--1481, Aug. 2009.

\bibitem{URV_THESIS}
U.~{Rau}, {\em Parameterized Deconvolution for Wideband Radio Synthesis
  Imaging}.
\newblock PhD thesis, New Mexico Institute of Mining and Technology, Socorro,
  NM, USA, May 2010.

\bibitem{CorMagFASR}
{Gary, D.E.}, ``Coronal magnetography with fasr,'' tech. rep., 2001.

\bibitem{Fleishman2009}
G.~D. {Fleishman}, G.~M. {Nita}, and D.~E. {Gary}, ``{Dynamic Magnetography of
  Solar Flaring Loops},'' {\em apjl}~{\bf 698}, pp.~L183--L187, June 2009.

\bibitem{Fleishman2010Concept}
G.~{Fleishman}, D.~{Gary}, G.~{Nita}, D.~{Alexander}, M.~{Aschwanden},
  T.~{Bastian}, H.~{Hudson}, G.~{Hurford}, E.~{Kontar}, D.~{Longcope},
  Z.~{Mikic}, M.~{DeRosa}, J.~{Ryan}, and S.~{White}, ``{Uncovering Mechanisms
  of Coronal Magnetism via Advanced 3D Modeling of Flares and Active
  Regions},'' {\em ArXiv e-prints} , Nov. 2010.

\bibitem{Fleishman_modeling2011}
G.~D. {Fleishman}, G.~M. {Nita}, and D.~E. {Gary}, ``{New interactive solar
  flare modeling and advanced radio diagnostics tools},'' in {\em IAU
  Symposium},  A.~{Bonanno}, E.~{de Gouveia Dal Pino}, and A.~G. {Kosovichev},
  eds., {\em IAU Symposium} {\bf 274}, pp.~280--283, June 2011.

\bibitem{FASRImaging2003}
S.~{White}, J.~{Lee}, M.~A. {Aschwanden}, and T.~S. {Bastian}, ``{Imaging
  capabilities of the Frequency Agile Solar Radiotelescope (FASR)},'' in {\em
  Society of Photo-Optical Instrumentation Engineers (SPIE) Conference Series},
   S.~L. {Keil} and S.~V. {Avakyan}, eds., {\em Society of Photo-Optical
  Instrumentation Engineers (SPIE) Conference Series} {\bf 4853}, pp.~531--541,
  Feb. 2003.

\end{thebibliography}
\bibliographystyle{spiebib}   

\end{document}